# Study of Citation Networks in Tribology Research

Bakthavachalam Elango
Library, IFET College of Engineering, Villupuram, Tamilnadu, India
elangokb@yahoo.com

Lutz Bornmann
Division for Science and Innovation Studies, Administrative Headquarters of the Max Planck Society, Hofgartenstr. 8, 80539 Munich, Germany
bornmann@gv.mpg.de

Subramaniam Shankar
Department of Mechatronics Engineering, Kongu Engineering College, Erode, Tamilnadu, India
shankariitm@gmail.com

**ABSTRACT**

CitNetExplorer has been used to study the citation networks among the scientific publications on tribology during the 15 years period from 1998-2012. Three data sets from Web of Science have been analyzed: (1) Core publications of tribology research, (2) publications on nanotribology and (3) publications of Bharat Bhushan (a top-contributor to nanotribology research). Based on this study, some suggestions are made to improve the CitNetExplorer.

## INTRODUCTION

Recently, van Eck & Waltman (2014a) introduced a new software tool called "CitNetExplorer" (abbreviation of Citation Network Explorer) for analyzing and visualizing the citation networks of scientific publications. The software is based on Garfield's work/ software on algorithmic historiography – HistCite (Garfield, Pudovkin & Istomin 2003; Leydesdroff 2010). Whereas HistCite is intended to analyze rather small publication sets, CitNetExplorer can handle millions of publications and citation relations. CitNetExplorer can be applied at the macro-level as well as the micro-level and offers the following functionality (van Eck & Waltman 2014b):

- Visualization of a citation network of a set of publications.
- Identification of related publications by identifying the connected components, clusters, or core sets in citation networks.
- Identification of predecessors (cited by selected publications) and successors (citing selected publications) of a publication set.
- Identification of publications located on a citation path between selected (two) publications. The publications on the path are referred to as intermediate publications.
- Drilling down into a citation network to reduce the number of selected publications (e.g. from total publications of tribology to publications on nanotribology).

The software can be applied for the following purposes (van Eck & Waltman 2014a):

- Studying the development of a research field over time: By showing the most important publications in a field, ordered by the year in which they appeared and the citation relations between these publications, one can inspect the development of a field over time.

- Studying the publication oeuvre of a researcher: The software shows the earlier literature on which a researcher builds his/ her work and the more recent literature that has been influenced by the work of a researcher.
- Literature reviewing: Obtaining an overview of a complex research topic (such as tribology) is a time consuming process as the publications of this research area appears in multiple scientific fields. CitNetExplorer simplifies systematic literature searches in various ways specifically by selecting all publications citing or cited by a given set of publications.

## OBJECTIVES

The aim of this study is to explore the potential of CitNetExplorer by using tribology research published during 1998-2012 and covered by Web of Science (WoS, Thomson Reuters) as an example.

## METHODS

### Dataset used

WoS was used to retrieve the bibliographic records related to tribology research. The following search query was used (Elango, Rajendran & Bornmann, submitted): **tribolog* OR "tribosyst*" OR "tribo-syst*" OR "tribo-chem*" OR "tribochem*" OR "tribotechn*" OR "tribo-physi*" OR "tribophysi*"*. The search yielded 15732 records during the period 1998-2012. The retrieved data has been extracted as tab-delimited text file.

### CitNetExplorer

In the citation network map (see e.g. Figure 1), each node represents a publication and each edge represents a citation relation between two publications. Edges are directed. They start at the citing publication (predecessor) and they end at the cited publication (successor). A publication either does or does not cite a certain other publication. A publication is labeled with the last name of the first author in the visualization. Each node has the following attributes along with standard bibliographic data such as authors, title and source.

- Publication year: The year in which a publication appeared.
- Citation score: The number of citations is calculated by two ways as internal and external. Internal citation score is the number of citations of a publication within a citation network being analyzed. External citation score considers outside citations. For example, internal citation score of a tribology publication equals the number of citations from other tribology publications. External citation score of a tribology publication equals the number of citations from all publications (in the WoS database).
- Group: Each publication can be algorithmically assigned to a group using clustering technique. Clustering is the division of data into groups of similar objects (Berkhin, 2006).

Two techniques can be used in CitNetExplorer for the following purposes: (1) first technique identifies core publications in a network and the other (2) clusters the publications in certain groups. A core publication in a citation network is defined as a publication that has at least a certain minimum number of citation relations (here we take 10 relations) with other core publications. Elango, Rajendran and Bornmann (2013) visualized the k-core of co-authorship network in nanotribology research output with 10 interlinks. The concept of core publications is based upon the k-cores introduced by Seidman (1983). (2) Clustering means that each publication in a citation network can be assigned to a cluster in such a way that those publications which are closer to each other, tend to be in the same cluster. So, each cluster consists of publications that are strongly connected in terms of citation relations. The main parameter in clustering is the resolution (Perbet, Stenger and Maki 2009) which controls the size of the generated clusters (see an example in table 1). As recommended by van Eck and Waltman (2014c), 0.75 is used as resolution parameter for clustering the tribology publications.

| **Table 1.** Resolution parameter vs. number of clusters | |
|---|---|
| Resolution parameter | Number of clusters |
| 0.75 | 3 |
| 1.00 | 4 |
| 2.00 | 8 |
| 3.00 | 9 |
| 4.00 | 10 |
| 5.00 | 13 |

**RESULTS**

The tribology research output published during 1998-2012 is analyzed as follows: (1) Core publications are identified among the tribology research output. (2) The publications on nanotribology are analyzed. (3) The publications of Bharat Bhushan – a top-contributor to nanotribology research are analyzed.

**Core publications in tribology**

A total of 15732 publications are involved in 54796 citation links between these publications during the study period. Table 2 provides an overview of citation links in three five-year block periods. Highest citation links and relative publications are observed in the second five-year period (2003 – 2007).

| **Table2**. Evolution of citation links | | |
|---|---|---|
| **Block Period** | **Citation Links** | **Publications** |
| 1998 – 2002 | 2467 | 3324 |
| 2003 – 2007 | 17281 | 8416 |
| 2008 – 2012 | 9407 | 7316 |

453 publications are determined as core publications by CitNetExplorer. After drilling down the core publications, clusters of closely related publications are identified. Out of 453 core publications, the 40 most frequently cited publications are displayed in Figure 1. Core publications are grouped into three clusters with 191 (42%) to group 1 (blue), 171 (38%) to group 2 (purple) and 91 (20%) to group 3 (green). The location of publications in horizontal position is determined by the closeness of the publications. Generally a citing publication is located below the corresponding cited publication. The bibliographic details of the 40 most frequently cited tribology publications are given in Table 3.

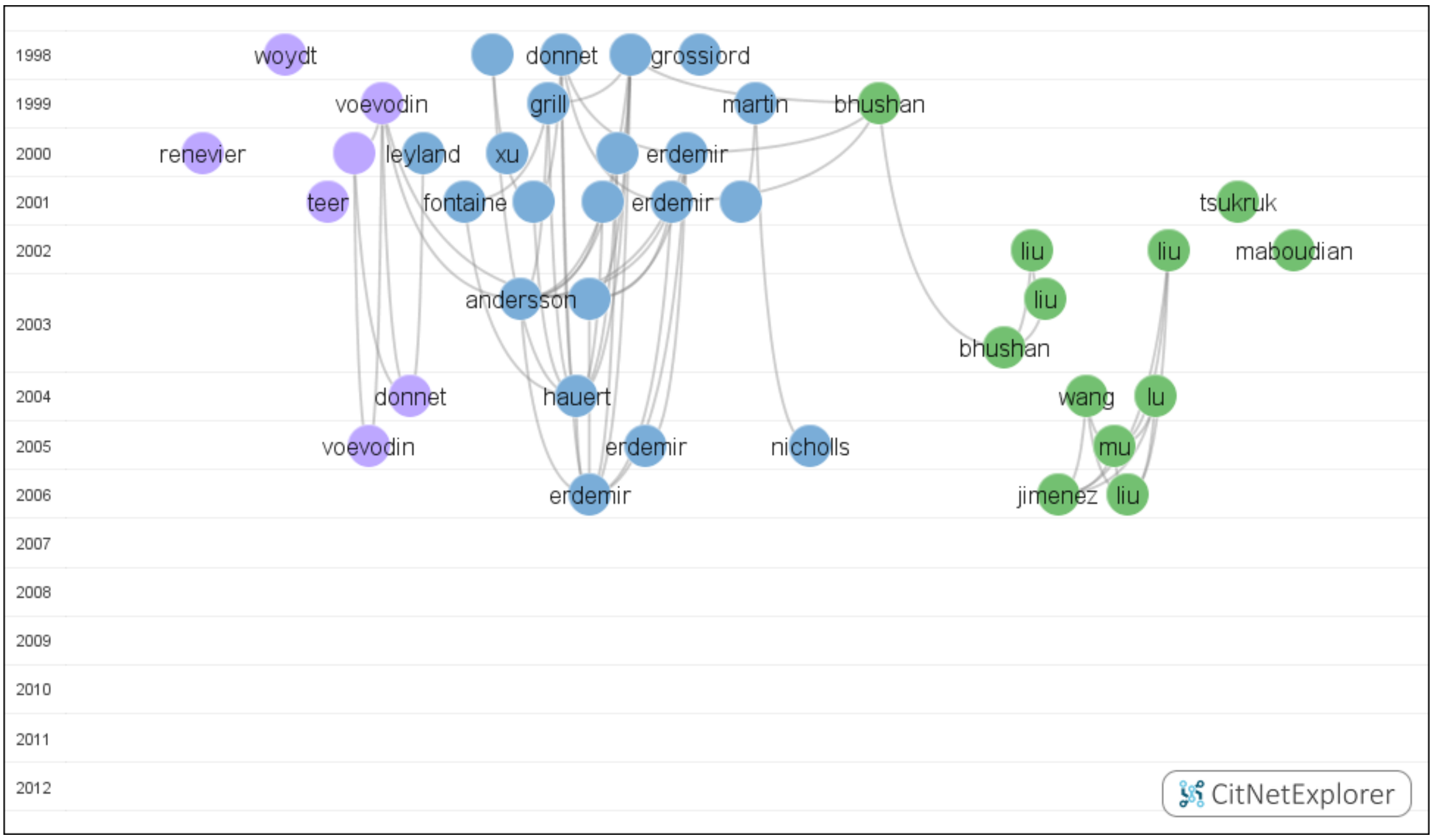

Figure 1 – Visualization of citation networks of core publications in tribology research during 1998-2012

As the table shows, the most cited publications deal with coating aspects over the substrate material to increase the overall hardness of the material.  In the visualization, the purple cluster covers the tribological publications which deal with advanced coating materials. These materials can be coated over the substrate to increase the wear resistance for various applications. The blue cluster covers the tribological publications particularly including the coating in which the specimens are tested with different lubricated conditions. Even though the material is coated, it needs to be tested for different working conditions to show its service life in practical conditions. The blue cluster publications are more or less similar to those in the purple cluster but the working conditions are tested in the studies. The green cluster covers the publications on nanotribology. In the publications of this cluster, some new phenomena like adhesion are tested for the coating conditions. Also, lubrication aspects which are used to reduce the friction and wear are considered. New solid, chemical and liquid lubricants are used and the consequences are tested for different substrate systems.

**Table 3**. Bibliographic details of the 40 most frequently cited publications

| Cluster | Article | Abbreviation |
|---|---|---|
| Purple | Woydt, M; Skopp, A; Dorfel, I; Witke, K (1998). Wear engineering oxides/anti-wear oxides. *Wear* | woydt |
| | Voevodin, AA; O'Neill, JP; Zabinski, JS (1999). Nanocompositetribological coatings for aerospace applications. *Surface and Coatings Technology* | voevodin |
| | Renevier, NM; Fox, VC; Teer, DG; Hampshire, J (2000). Coating characteristics and tribological properties of sputter-deposited MoS2/metal composite coatings deposited by closed field unbalanced magnetron sputter ion plating. *Surface and Coatings Technology* | renevier |
| | Voevodin, AA; Zabinski, JS (2000). Supertough wear-resistant coatings with 'chameleon' surface adaptation. *Thin Solid Films* | voevodin |
| | Teer, DG (2001). New solid lubricant coatings. *Wear* | teer |
| | Donnet, C; Erdemir, A (2004). Historical developments and new trends in tribological and solid lubricant coatings. *Surface and Coatings Technology* | donnet |
| | Voevodin, AA; Zabinski, JS (2005). Nanocomposite and nanostructured tribological materials for space applications. *Composites Science and Technology* | voevodin |
| Blue | Ronkainen, H; Varjus, S; Holmberg, K (1998). Friction and wear properties in dry, water- and oil-lubricated DLC against alumina and DLC against steel contacts. *Wear* | ronkainen |
| | Donnet, C (1998). Recent progress on the tribology of doped diamond-like and carbon alloy coatings: a review. *Surface and Coatings Technology* | donnet |
| | Donnet, C; Le Mogne, T; Ponsonnet, L; Belin, M; Grill, A; Patel, V; Jahnes, C (1998). The respective role of oxygen and water vapor on the tribology of hydrogenated diamond-like carbon coatings. *Tribology Letters* | donnet |
| | Grossiord, C; Varlot, K; Martin, JM; Le Mogne, T; Esnouf, C; Inoue, K (1998). MoS2, single sheet lubrication by molybdenum dithioca. *Tribology International* | grossiord |
| | Grill, A (1999). Diamond-like carbon: state of the art. *Diamond and Related Materials* | grill |
| | Martin, JM (1999). Antiwear mechanisms of zinc dithiophosphate: a chemical hardness approach. *Tribology Letters* | martin |
| | Leyland, A; Matthews, A (2000). On the significance of the H/E ratio in wear control: a nanocomposite coating approach to optimized tribological behavior. *Wear* | leyland |
| | Xu, JG; Kato, K (2000). Formation of tribochemical layer of ceramics sliding in water and its role for low friction. *Wear* | xu |
| | Erdemir, A; Eryilmaz, OL; Nilufer, IB; Fenske, GR (2000). Synthesis of superlow-friction carbon films from highly hydrogenated methane plasmas. *Surface and Coatings Technology* | erdemir |
| | Erdemir, A; Eryilmaz, OL; Fenske, G (2000). Synthesis of diamondlike carbon films with superlow friction and wear properties. *Journal of Vaccum Science and Technology A – Vacuum Surfaces and Films* | erdemir |
| | Fontaine, J; Donnet, C; Grill, A; LeMogne, T (2001). Tribochemistry between hydrogen and diamond-like carbon films. *Surface and Coatings Technology* | fontaine |
| | Ronkainen, H; Varjus, S; Holmberg, K (2001). Tribological performance of different DLC coatings in water-lubricated conditions. *Wear* | ronkainen |
| | Ronkainen, H; Varjus, S; Koskinen, J; Holmberg, K (2001). Differentiating the tribological performance of hydrogenated and hydrogen-free DLC coatings. *Wear* | ronkainen |
| | Erdemir, A (2001). The role of hydrogen in tribological properties of diamond-like carbon films. *Surface and Coatings Technology* | erdemir |
| | Martin, JM; Grossiord, C; Le Mogne, T; Bec, S; Tonck, A (2001). The two-layer structure of zndtptribofilms Part 1: AES, XPS and XANES analyses. *Tribology International* | martin |
| | Andersson, J; Erck, RA; Erdemir, A (2003). Friction of diamond-like carbon films | andersson |
| | in different atmospheres. *Wear* | |
| | Andersson, J; Erck, RA; Erdemir, A (2003). Frictional behavior of diamondlike carbon films in vacuum and under varying water vapor pressure. *Surface and Coatings Technology* | andersson |
| | Hauert, R (2004). An overview on the tribological behavior of diamond-like carbon in technical and medical applications. *Tribology International* | hauert |
| | Erdemir, A (2005). Review of engineered tribological interfaces for improved boundary lubrication. *Tribology International* | erdemir |
| | Nicholls, MA; Do, T; Norton, PR; Kasrai, M; Bancroft, GM (2005). Review of the lubrication of metallic surfaces by zinc dialkyl-dithlophosphates. *Tribology International* | nicholls |
| | Erdemir, A; Donnet, C (2006). Tribology of diamond-like carbon films: recent progress and future prospects. *Journal of Physics D – Applied Physics* | erdemir |
| Green | Bhushan, B (1999). Chemical, mechanical and tribological characterization of ultra-thin and hard amorphous carbon coatings as thin as 3.5 nm: recent developments. *Diamond and Related Materials* | bhushan |
| | Tsukruk, VV (2001). Molecular lubricants and glues for micro- and nanodevices. *Advanced Materials* | tsukuk |
| | Liu, H; Bhushan, B (2002). Investigation of nanotribological properties of self-assembled monolayers with alkyl and biphenyl spacer chains. *Ultramicroscopy* | liu |
| | Liu, WM; Ye, CF; Gong, QY; Wang, HZ; Wang, P (2002). Tribological performance of room-temperature ionic liquids as lubricant. *Tribology Letters* | liu |
| | Maboudian, R; Ashurst, WR; Carraro, C (2002). Tribological challenges in micromechanical systems. *Tribology Letters* | maboudian |
| | Bhushan, B (2003). Adhesion and stiction: Mechanisms, measurement techniques, and methods for reduction. *Journal of Vacuum Science and Technology B* | bhushan |
| | Liu, HW; Bhushan, B (2003). Nanotribological characterization of molecularly thick lubricant films for applications to MEMS/NEMS by AFM. *Ultramicroscopy* | liu |
| | Wang, HZ; Lu, QM; Ye, CF; Liu, WM; Cui, ZJ (2004). Friction and wear behaviors of ionic liquid of alkylimidazoliumhexafluorophosphates as lubricants for steel/steel contact. *Wear* | wang |
| | Lu, QM; Wang, HZ; Ye, CF; Liu, WM; Xue, QJ (2004). Room temperature ionic liquid 1-ethyl-3-hexylimidazolium-bis (trifluoromethylsulfonyl)-imide as lubricant for steel-steel contact. *Tribology International* | lu |
| | Mu, ZG; Zhou, F; Zhang, SX; Liang, YM; Liu, WM (2005). Effect of the functional groups in ionic liquid molecules on the friction and wear behavior of aluminum alloy in lubricated aluminum-on-steel contact. *Tribology International* | mu |
| | Jimenez, AE; Bermudez, MD; Iglesias, P; Carrion, FJ; Martinez-Nicolas, G (2006). 1-N-alkyl-3-methylimidazolium ionic liquids as neat lubricants and lubricant additives in steel-aluminium contacts. *Wear* | jimenez |
| | Liu, XQ; Zhou, F; Liang, YM; Liu, WM (2006). Tribological performance of phosphonium based ionic liquids for an aluminum-on-steel system and opinions on lubrication mechanism. *Wear* | liu |

**Analysis of the publications on a specific topic: nanotribology**

From the full network, we drill down a sub network consisting of publications on nanotribology. To extract the nanotribology publications, we use the following keywords (Elango, Rajendran & Bornmann 2013) in title: **nanotri* OR *microtri* OR *nano-tri* OR *micro-tri* OR *nano tri* OR *micro tri**. Initially 379 publications are identified. We obtain a citation network of nanotribology publications <u>without</u> intermediate publications (a second analysis including intermediate publications is described below). Intermediate publications are those in a network that are located on a citation path between two marked publications. Clustering is done using this publication set. Five clusters appear where by 178 publications do not belong to any of the five clusters (see Figure 2). Of the total of 379 publications, 40 most frequently cited publications are displayed in the visualization. Bibliographic details of these publications are provided in Table 4. Publications which are not included in any of the clusters in the figure are displayed with grey color.

It is hard to differentiate the publications in the clusters of Figure 2 in terms of content. In the visualization, most of the publications authored by Bharat Bhushan are grouped under the blue cluster. It seems that a cluster may represent the publications of an author rather than those of a specific topic (see here: van Eck and Waltman 2014a). The purple cluster covers several publications related to tribochemistry and molecular dynamics. These studies compare the micro level tribological properties and its advantages of coated over pure samples. Also, the papers report the advancement of nanotribological characteristics and properties, adhesive mechanism and asperity based contact simulations. The green cluster is determined by publications which are related to macro and nanotribology, coating depth measurement using atomic force microscopical concepts, influence of coating over the substrate, the importance of carbon based films and some further topics. The studies are based on using advanced instruments like Atomic Force Microscopy (AFM) in nanotribological investigations. Few papers emphasize the importance of using carbon based nano/micro films over the surface to improve the substrate hardness. The publication of Sundararajan & Bhushan (1998) dealing with Micro-electro-mechanical systems (MEMS) and located in the orange cluster is linked to the green and purple clusters. Similar to the papers in the green and purple clusters, coating and the usage of Atomic Force Microscopy (AFM) in tribological study is explored in both orange cluster papers. The "santos" publication in the orange cluster is isolated as it deals with diamond-like carbon (DLC). There is a mix of very different publications in the grey cluster covering the topics of e.g. carbon, molecular dynamics, and nanotubes. One common property of the grey cluster publications mix is the missing link to other publications in the visualization. There is an isolated publication colored yellow covering the topic of thin films.

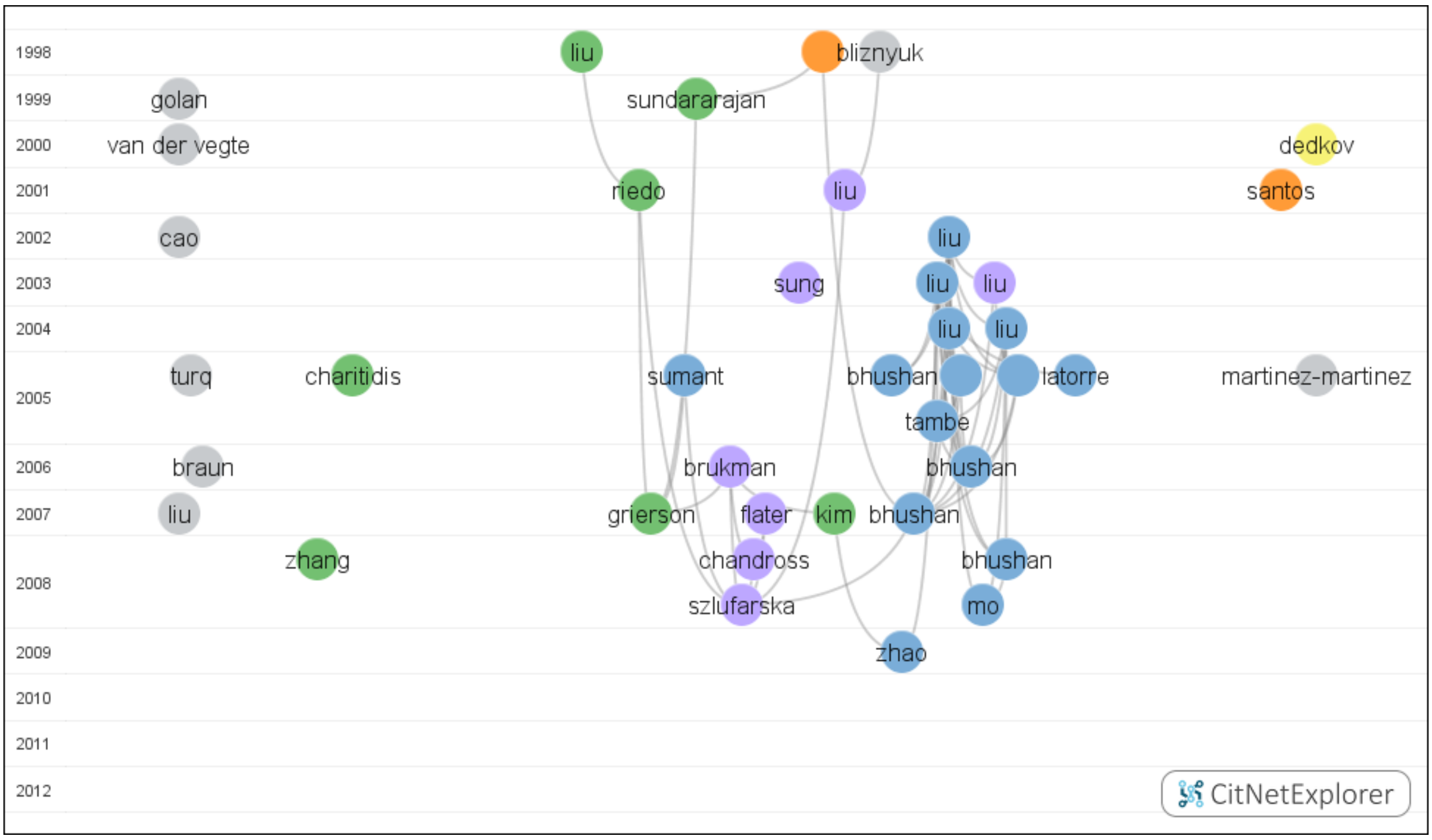

Figure 2 – Visualization of citation network of nanotribology publications without intermediate publications

| Table 4. Bibliographic details of the 40 nanotribology publications (without intermediate publications) | | |
|---|---|---|
| **Cluster** | **Article** | **Abbreviation** |
| Green | Liu, E; Blanpain, B; Celis, JP; Roos, JR (1998). Comparative study between macrotribology and nanotribology. *Journal of Applied Physics* | liu |
| | Sundararajan, S; Bhushan, B (1999). Micro/nanotribology of ultra-thin hard amorphous carbon coatings using atomic force friction force microscopy. *Wear* | sundararajan |
| | Riedo, E; Chevrier, J; Comin, F; Brune, H (2001). Nanotribology of carbon based thin films: the influence of film structure and surface morphology. *Surface Science* | riedo |
| | Charitidis, C; Logothetidis, S (2005). Nanomechanical and nanotribological properties of carbon based films. *Thin Solid Films* | charitidis |
| | Grierson, DS; Carpick, RW (2007). Nanotribology of carbon-based materials. *Nano Today* | grierson |
| | Kim, SH; Asay, DB; Dugger, MT (2007). Nanotribology and MEMS. *Nano Today* | kim |
| | Zhang, HS; Endrino, JL; Anders, A (2008). Comparative surface and nano-tribological characteristics of nanocomposite diamond-like carbon thin films doped by silver. *Applied Surface Science* | zhang |
| Purple | Liu, HW; Ahmed, SIU; Scherge, M (2001). Microtribological properties of silicon and silicon coated with diamond like carbon, octadecyltrichlorosilane and stearic acid cadmium salt films: A comparative study. *Thin Solid Films* | liu |
| | Sung, IH; Yang, JC; Kim, DE; Shin, BS (2003). Micro/nano-tribological characteristics of self-assembled monolayer and its application in nano-structure fabrication. *Wear* | sung |
| | Liu, HW; Bhushan, B (2003). Adhesion and friction studies of microelectromechanical systems/nanoelectromechanical systems materials using a novel microtriboapparatus. *Journal of Vacuum Science and Technology A* | liu |
| | Brukman, MJ; Marco, GO; Dunbar, TD; Boardman, LD; Carpick, RW (2006). Nanotribological properties of alkanephosphonic acid self-assembled monolayers on aluminum oxide: Effects of fluorination and substrate crystallinity. *Langmuir* | brukman |
| | Flater, EE; Ashurst, WR; Carpick, RW (2007). Nanotribology of octadecyltrichlorosilane monolayers and silicon: Self-mated versus unmated interfaces and local packing density effects. *Langmuir* | flater |
| | Chandross, M; Lorenz, CD; Stevens, MJ; Grest, GS (2008). Simulations of nanotribology with realistic probe tip models. *Langmuir* | chandross |
| | Szlufarska, I; Chandross, M; Carpick, RW (2008). Recent advances in single-asperity nanotribology. *Journal of Physics D – Applied Physics* | szlufarska |
| Blue | Liu, H; Bhushan, B (2002). Investigation of nanotribological properties of self-assembled monolayers with alkyl and biphenyl spacer chains (Invited). *Ultramicroscopy* | liu |
| | Liu, HW; Bhushan, B (2003). Nanotribological characterization of molecularly thick lubricant films for applications to MEMS/NEMS by AFM. *Ultramicroscopy* | liu |
| | Liu, HW; Bhushan, B (2004). Nanotribological characterization of digital micromirror devices using an atomic force microscope. *Ultramicroscopy* | liu |
| | Liu, HW; Bhushan, B (2004). Investigation of nanotribological and nanomechanical properties of the digital micromirror device by atomic force microscopy. *Journal of Vacuum Science and Technology A* | liu |
| | Sumant, AV; Grierson, DS; Gerbi, JE; Birrell, J; Lanke, UD; Auciello, O; Carlisle, JA; Carpick, RW (2005). Toward the ultimate tribological interface: Surface chemistry and nanotribology of ultrananocrystalline diamond. *Advanced Materials* | sumant |
| | Bhushan, B (2005). Nanotribology and nanomechanics. *Wear* | bhushan |
| | Kasai, T; Bhushan, B; Kulik, G; Barbieri, L; Hoffmann, P (2005). Micro/nanotribological study of perfluorosilane SAMs for antistiction and low wear. *Journal of Vacuum Science and Technology B* | kasai |
| | Lee, KK; Bhushan, B; Hansford, D (2005). Nanotribological characterization of fluoropolymer thin films for biomedical micro/nanoelectromechanical system applications. *Journal of Vacuum Science and Technology A* | lee |
| | LaTorre, C; Bhushan, B (2005). Nanotribological characterization of human hair and skin using atomic force microscopy. *Ultramicroscopy* | latorre |
| | Tambe, NS; Bhushan, B (2005). Nanotribological characterization of self-assembled monolayers deposited on silicon and aluminium substrates. *Nanotechnology* | tambe |
| | Bhushan, B; Cichomski, M; Hoque, E; DeRose, JA; Hoffmann, P; Mathieu, HJ (2006). Nanotribological characterization of perfluoroalkylphosphonate self-assembled monolayers deposited on aluminum-coated silicon substrates. *Microsystem Technologies –Micro-and Nanosystems-Information Storage and Processing Systems* | bhushan |
| | Bhushan, B (2007). Nanotribology and nanomechanics of MEMS/NEMS and BioMEMS/BioNEMS materials and devices. *Microelectronic Engineering* | bhushan |
| | Bhushan, B; Palacio, M; Kinzig, B (2008). AFM-based nanotribological and electrical characterization of ultrathin wear-resistant ionic liquid films. *Journal of Colloid and Interface Science* | bhushan |
| | Mo, YF; Zhao, WJ; Zhu, M; Bai, MW (2008). Nano/Microtribological Properties of Ultrathin Functionalized Imidazolium Wear-Resistant Ionic Liquid | mo |
| | Films on Single Crystal Silicon. *Tribology Letters* | |
| | Zhao, WJ; Zhu, M; Mo, YF; Bai, MW (2009). Effect of anion on micro/nano-tribological properties of ultra-thin imidazolium ionic liquid films on silicon wafer. *Colloids and Surfaces A – Physicochemical and Engineering Aspects* | zhao |
| Orange | Sundararajan, S; Bhushan, B (1998). Micro/nanotribological studies of polysilicon and SiC films for MEMS applications. *Wear* | sundararajan |
| | Santos, LV; Trava-Airoldi, VJ; Iha, K; Corat, EJ; Salvadori, MC (2001). Diamond-like-carbon and molybdenum disulfide nanotribology studies using atomic force measurements. *Diamond and Related Materials* | santos |
| Yellow | Dedkov, GV (2000). Experimental and theoretical aspects of the modern nanotribology. *Physica Status Solidi A – Applications and Materials Science* | dedkov |
| Not clustered / Grey | Bliznyuk, VN; Everson, MP; Tsukruk, VV (1998). Nanotribological properties of organic boundary lubricants: Langmuir films versus self-assembled monolayers. *Journal of Tribology – Transactions of the ASME* | bliznyuk |
| | Golan, Y; Drummond, C; Homyonfer, M; Feldman, Y; Tenne, R; Israelachvili, J (1999). Microtribology and direct force measurement of WS2 nested fullerene-like nanostructures. *Advanced Materials* | golan |
| | van der Vegte, EW; Subbotin, A; Hadziioannou, G; Ashton, PR; Preece, JA (2000). Nanotribological properties of unsymmetrical n-dialkyl sulfide monolayers on gold: Effect of chain length on adhesion, friction, and imaging. *Langmuir* | ven der Vegte |
| | Cao, TB; Wei, F; Yang, YL; Huang, L; Zhao, XS; Cao, WX (2002). Microtribologic properties of a covalently attached nanostructured self-assembly film fabricated from fullerene carboxylic acid and diazoresin. *Langmuir* | cao |
| | Turq, V; Ohmae, N; Martin, JM; Fontaine, J; Kinoshita, H; Loubet, J (2005). Influence of humidity on microtribology of vertically aligned carbon nanotube film. *Tribology Letters* | turq |
| | Martinez-Martinez, D; Sanchez-Lopez, JC; Rojas, TC; Fernandez, A; Eaton, P; Belin, M (2005). Structural and microtribological studies of Ti-C-N based nanocomposite coatings prepared by reactive sputtering. *Thin Solid Films* | martinez-martinez |
| | Braun, OM; Naumovets, AG (2006). Nanotribology: Microscopic mechanisms of friction. *Surface Science Reports* | braun |
| | Liu, LN; Gu, AJ; Fang, ZP; Tong, LF; Xu, ZB (2007). The effects of the variations of carbon nanotubes on the micro-tribological behavior of carbon nanotubes/bismaleimidenanocomposite. *Composites Part A – Applied Science and Manufacturing* | liu |

In a second analysis of the publications dealing with the topic of nanotribology, we obtain a citation network of nanotribology publications with intermediate publications. After drilling down with intermediate publications, we obtain a citation network consisting of 1065 publications which are clustered. We receive 7 clusters; whereby 105 publications do not belong to any of the seven clusters (see Figure 3). The most 40 frequently cited publications are displayed in the visualization and the bibliographic details of these publications are provided in Table 5.

In the network of Figure 3, most of the nanotribology publications are grouped into the blue colored cluster. Most of these papers are related to coating and its effective applications. They report the synthesis of advanced coating materials, usage of Atomic Force Microscopy (AFM), characterization of the materials, and applications relating to coating. The green cluster covers the publications including basic comparative studies between macro and nanotribology. Studies relating to lubricating aspects include hydrogen, nitrogen, water vapor, and different atmospheric conditions which are utilized in the papers. All the papers in the green cluster either deal with coatings or lubricating aspects. The purple cluster covers publications on thin films which are made of composites. Nano composite coatings reported in the publications of this cluster are effectively utilized to reduce friction and wear. Specific applications of composite coatings which are related to space design are also reported. The orange cluster contains publications which deal with e.g. molecular dynamics and fullerene. The pink colored publication in Figure 3 deals with nanotribology, specifically deformation, and asperity.

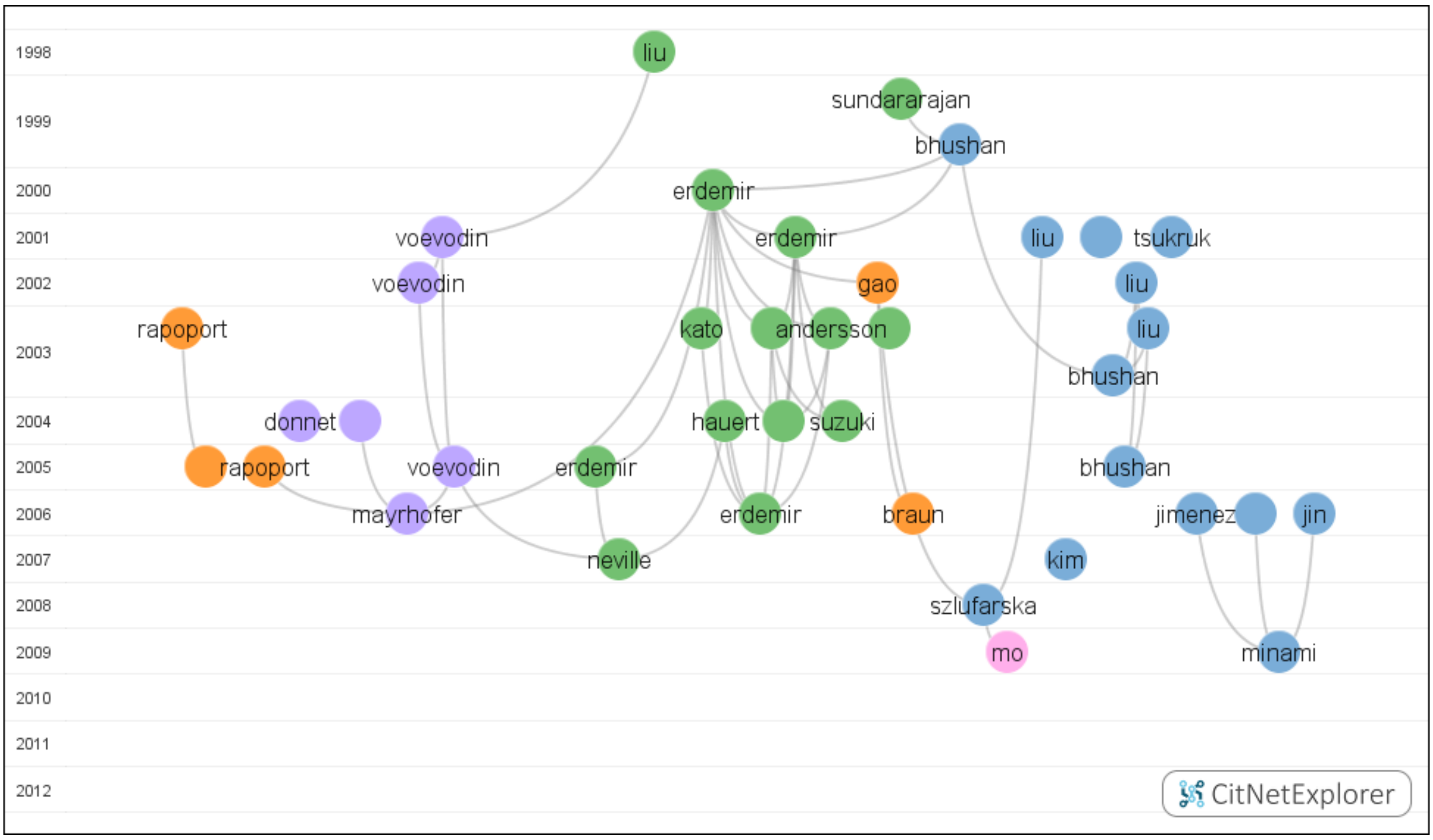

Figure 3 – Visualization of citation network of nanotribology publications with intermediate publications

| **Table 5**. Bibliographic details of the 40 nanotribology publications (with intermediate publications) | | |
|---|---|---|
| **Cluster** | **Article** | **Abbreviation** |
| Orange | Rapoport, L; Leshchinsky, V; Lapsker, I; Volovik, Y; Nepomnyashchy, O; | rapoport |
| | Lvovsky, M; Popovitz-Biro, R; Feldman, Y; Tenne, R (2003). Tribological properties of WS2 nanoparticles under mixed lubrication. *Wear* | |
| | Joly-Pottuz, L; Dassenoy, F; Belin, M; Vacher, B; Martin, JM; Fleischer, N (2005). Ultralow-friction and wear properties of IF-WS2 under boundary lubrication. *Tribology Letters* | joly-pottuz |
| | Rapoport, L; Fleischer, N; Tenne, R (2005). Applications of WS2 (MoS2) inorganic nanotubes and fullerene-like nanoparticles for solid lubrication and for structural nanocomposites. *Journal of Materials Chemistry* | rapoport |
| | Gao, GT; Mikulski, PT; Harrison, JA (2002). Molecular-scale tribology of amorphous carbon coatings: Effects of film thickness, adhesion, and long-range interactions. *Journal of the American Chemical Society* | gao |
| | Braun, OM; Naumovets, AG (2006). Nanotribology: Microscopic mechanisms of friction. *Surface Science Reports* | braun |
| Purple | Voevodin, AA; Hu, JJ; Fitz, TA; Zabinski, JS (2001). Tribological properties of adaptive nanocomposite coatings made of yttria stabilized zirconia and gold. *Surface and Coatings Technology* | voevodin |
| | Voevodin, AA; Fitz, TA; Hu, JJ; Zabinski, JS (2002). Nanocomposite tribological coatings with "chameleon" surface adaptation. *Journal of Vacuum Science and Technology A – Vacuum Surfaces and Films* | voevodin |
| | Donnet, C; Erdemir, A (2004). Historical developments and new trends in tribological and solid lubricant coatings. *Surface and Coatings Technology* | donnet |
| | Mayrhofer, PH; Hovsepian, PE; Mitterer, C; Munz, WD (2004). Calorimetric evidence for frictional self-adaptation of TiAlN/VN superlattice coatings. *Surface and Coatings Technology* | mayrhofer |
| | Voevodin, AA; Zabinski, JS (2005). Nanocomposite and nanostructured tribological materials for space applications. *Composites Science and Technology* | voevodin |
| | Mayrhofer, PH; Mitterer, C; Hultman, L; Clemens, H (2006). Microstructural design of hard coatings. *Progress in Materials Science* | mayrhofer |
| Green | Liu, E; Blanpain, B; Celis, JP; Roos, JR (1998). Comparative study between macrotribology and nanotribology. *Journal of Applied Physics* | liu |
| | Sundararajan, S; Bhushan, B (1999). Micro/nanotribology of ultra-thin hard amorphous carbon coatings using atomic force friction force microscopy. *Wear* | sundararajan |
| | Erdemir, A; Eryilmaz, OL; Fenske, G (2000). Synthesis of diamondlike carbon films with superlow friction and wear properties. *Journal of Vacuum Science and Technology A – Vacuum Surfaces and Films* | erdemir |
| | Erdemir, A (2001). The role of hydrogen in tribological properties of diamond-like carbon films. *Surface and Coatings Technology* | erdemir |
| | Kato, K; Umehara, N; Adachi, K (2003). Friction, wear and N-2-lubrication of carbon nitride coatings: a review. *Wear* | kato |
| | Andersson, J; Erck, RA; Erdemir, A (2003). Frictional behavior of diamondlike carbon films in vacuum and under varying water vapor pressure. *Surface and Coatings Technology* | andersson |
| | Andersson, J; Erck, RA; Erdemir, A (2003). Friction of diamond-like carbon films in different atmospheres. *Wear* | andersson |
| | Jiang, JR; Zhang, S; Arnell, RD (2003). The effect of relative humidity on wear of a diamond-like carbon coating. *Surface and Coatings Technology* | jiang |
| | Hauert, R (2004). An overview on the tribological behavior of diamond-like carbon in technical and medical applications. *Tribology International* | hauert |
| | Erdemir, A (2004). Genesis of superlow friction and wear in diamondlike carbon films. *Tribology International* | erdemir |
| | Suzuki, M; Ohana, T; Tanaka, A (2004). Tribological properties of DLC films with different hydrogen contents in water environment. *Diamond and Related Materials* | suzuki |
| | Erdemir, A (2005). Review of engineered tribological interfaces for improved boundary lubrication. *Tribology International* | erdemir |
| | Erdemir, A; Donnet, C (2006). Tribology of diamond-like carbon films: recent progress and future prospects. *Journal of Physics D – Applied Physics* | erdemir |
| | Neville, A; Morina, A; Haque, T; Voong, Q (2007). Compatibility between tribological surfaces and lubricant additives - How friction and wear reduction can | neville |
| | be controlled by surface/lube synergies. *Tribology International* | |
| Blue | Bhushan, B (1999). Chemical, mechanical and tribological characterization of ultra-thin and hard amorphous carbon coatings as thin as 3.5 nm: recent developments. *Diamond and Related Materials* | bhushan |
| | Liu, HW; Ahmed, SIU; Scherge, M (2001). **Microtribological** properties of silicon and silicon coated with diamond like carbon, octadecyltrichlorosilane and stearic acid cadmium salt films: A comparative study. *Thin Solid Films* | liu |
| | Brewer, NJ; Beake, BD; Leggett, GJ (2001). Friction force microscopy of self-assembled monolayers: Influence of adsorbate alkyl chain length, terminal group chemistry, and scan velocity. *Langmuir* | brewer |
| | Tsukruk, VV (2001). Molecular lubricants and glues for micro- and nanodevices. *Advanced Materials* | tsukruk |
| | Liu, H; Bhushan, B (2002). Investigation of **nanotribological** properties of self-assembled monolayers with alkyl and biphenyl spacer chains (Invited). *Ultramicroscopy* | liu |
| | Liu, HW; Bhushan, B (2003). **Nanotribological** characterization of molecularly thick lubricant films for applications to MEMS/NEMS by AFM. *Ultramicroscopy* | liu |
| | Bhushan, B (2003). Adhesion and stiction: Mechanisms, measurement techniques, and methods for reduction. *Journal of Vacuum Science and Technology B* | bhushan |
| | Bhushan, B (2005). **Nanotribology** and nanomechanics. *Wear* | bhushan |
| | Jimenez, AE; Bermudez, MD; Iglesias, P; Carrion, FJ; Martinez-Nicolas, G (2006). 1-N-alkyl-3-methylimidazolium ionic liquids as neat lubricants and lubricant additives in steel-aluminium contacts. *Wear* | jimenez |
| | Qu, J; Truhan, JJ; Dai, S; Luo, H; Blau, PJ (2006). Ionic liquids with ammonium cations as lubricants or additives. *Tribology Letters* | qu |
| | Jin, CM; Ye, CF; Phillips, BS; Zabinski, JS; Liu, XQ; Liu, WM; Shreeve, JM (2006). Polyethylene glycol functionalized dicationic ionic liquids with alkyl or polyfluoroalkyl substituents as high temperature lubricants. *Journal of Materials Chemistry* | jin |
| | Kim, SH; Asay, DB; Dugger, MT (2007). **Nanotribology** and MEMS. *Nano today* | kim |
| | Szlufarska, I; Chandross, M; Carpick, RW (2008). Recent advances in single-asperity **nanotribology**. *Journal of Physics D – Applied Physics* | szlufarska |
| | Minami, I (2009). Ionic Liquids in Tribology. *Molecules* | minami |
| Pink | Mo, YF; Turner, KT; Szlufarska, I (2009). Friction laws at the nanoscale. *Nature* | mo |

## Analysis of Bharat Bhushan's publications

According to the personal biodata, Bharat Bhushan published more than 800 papers. He has published fundamental studies in the interdisciplinary areas of bio- / nanotribology / nanomechanics and nanomaterials characterization in bio- / nanotechnology and biomimetics with a focus on scanning probe techniques. He was the top contributor in nanotribology research during 1996-2010 (Elango, Rajendran & Bornmann 2013). We identified 158 publications of Bharat Bhushan in WoS using the following author keyword: bhushan*. We drill down the sub network with intermediate publications and obtain a citation network with 356 publications along with 1370 citation relations. These publications are clustered into 6 groups, whereby 11 publications do not belong to any cluster. Among the six clusters, the publications of four clusters are displayed in figure 4 (the 40 most frequently cited publications belong to these 4 clusters only). The bibliographic details of the 40 publications are given in Table 6. The publications of Bharat Bhushan are grouped into the blue cluster (with one exception); the other publications are intermediate publications.

Figure 4 shows Bharat Bhushan publications especially deal with surface texturing. This concept is effectively utilized to reduce wear and improved lubrication aspects without any specific coatings. The green cluster covers the publications on e.g. ionic liquids, micro-electro-mechanical systems (MEMS), and thin films. Adhesion, sliding contact consequences against different substrates, lubricating environments, and coatings using different deposition techniques are reported. Publications in the purple cluster cover topics as molecular dynamics and chemical deposition. Diamond-like carbon (DLC) is one such hard coating which gains importance in many purple cluster publications. Also, the single purple cluster publication of Bharat Bhushan covers this topic.

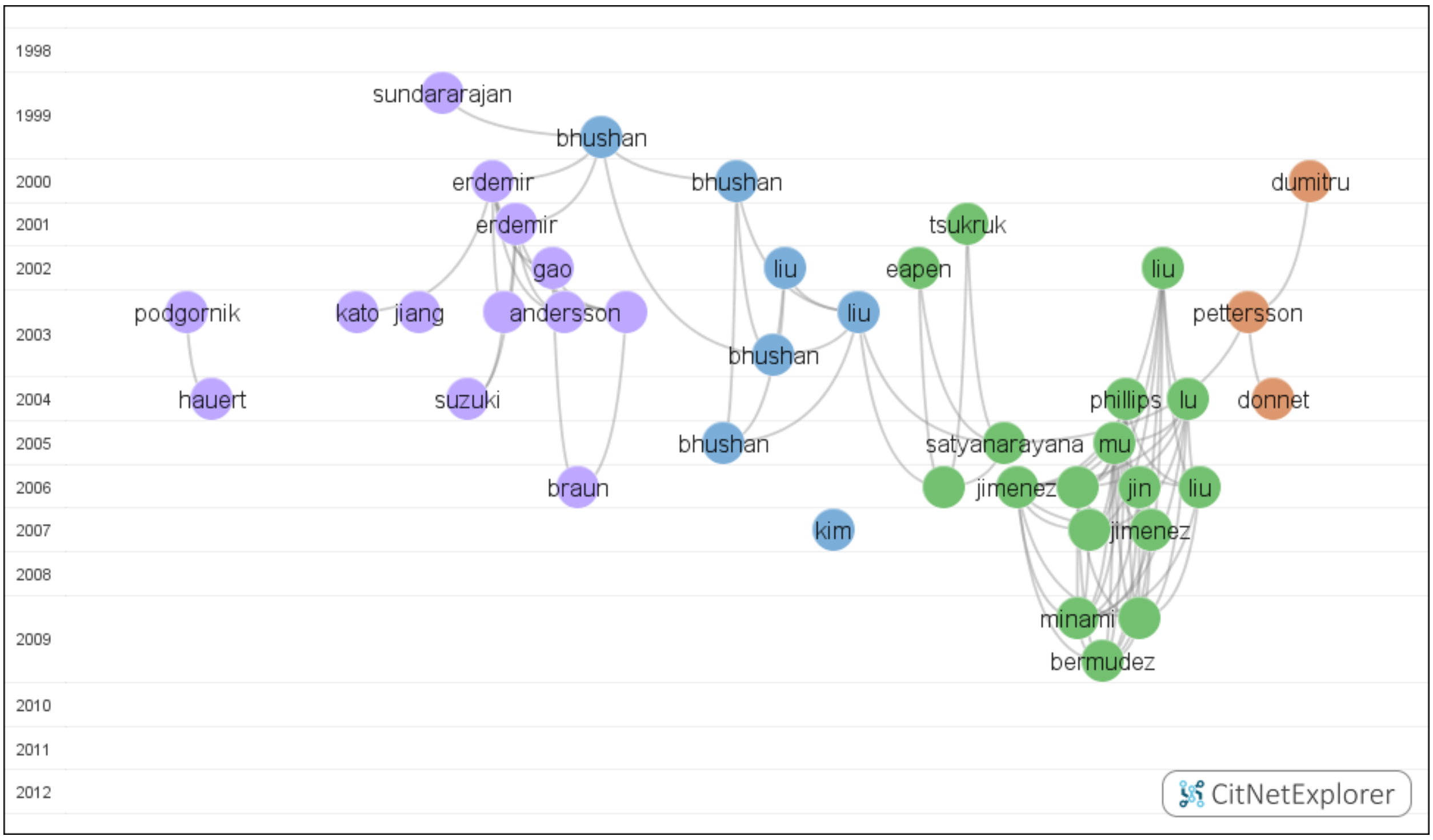

Figure 4 – Visualization of citation network of Bharat Bhushan with intermediate publications

| **Table 6**. Bibliographic details of the 40 most frequently cited publications by Bharat Bhushan | | |
|---|---|---|
| **Cluster** | **Article** | **Abbreviation** |
| Purple | Sundararajan, S; **Bhushan**, B (1999). Micro/nanotribology of ultra-thin hard amorphous carbon coatings using atomic force friction force microscopy. *Wear* | sundararajan |
| | Erdemir, A; Eryilmaz, OL; Fenske, G (2000). Synthesis of diamondlike carbon films with superlow friction and wear properties. *Journal of Vacuum Science and Technology D – Vacuum Surfaces and Films* | erdemir |
| | Erdemir, A (2001). The role of hydrogen in tribological properties of diamond-like carbon films. *Surface and Coatings Technology* | erdemir |
| | Gao, GT; Mikulski, PT; Harrison, JA (2002). Molecular-scale tribology of amorphous carbon coatings: Effects of film thickness, adhesion, and long-range interactions. *Journal of the American Chemical Society* | gao |
| | Podgornik, B; Jacobson, S; Hogmark, S (2003). DLC coating of boundary lubricated components - advantages of coating one of the contact surfaces rather than both or none. *Tribology International* | podgornik |
| | Kato, K; Umehara, N; Adachi, K (2003). Friction, wear and N-2-lubrication of carbon nitride coatings: a review. *Wear* | kato |
| | Jiang, JR; Zhang, S; Arnell, RD (2003). The effect of relative humidity on wear of a diamond-like carbon coating. *Surface and Coatings Technology* | jiang |
| | Andersson, J; Erck, RA; Erdemir, A (2003). Frictional behavior of diamondlike carbon films in vacuum and under varying water vapor pressure. *Surface and Coatings Technology* | andersson |
| | Andersson, J; Erck, RA; Erdemir, A (2003). Friction of diamond-like carbon films in different atmospheres. *Wear* | andersson |
| | Gao, GT; Mikulski, PT; Chateauneuf, GM; Harrison, JA (2003). The effects of film structure and surface hydrogen on the properties of amorphous carbon films. *Journal of Physical Chemistry B* | gao |
| | Hauert, R (2004). An overview on the tribological behavior of diamond-like carbon in technical and medical applications. *Tribology International* | hauert |
| | Suzuki, M; Ohana, T; Tanaka, A (2004). Tribological properties of DLC films with different hydrogen contents in water environment. *Diamond and Related Materials* | suzuki |
| | Braun, OM; Naumovets, AG (2006). Nanotribology: Microscopic mechanisms of friction. *Surface Science Reports* | braun |
| Blue | **Bhushan**, B (1999). Chemical, mechanical and tribological characterization of ultra-thin and hard amorphous carbon coatings as thin as 3.5 nm: recent developments. *Diamond and Related Materials* | bhushan |
| | **Bhushan**, B; Dandavate, C (2000). Thin-film friction and adhesion studies using atomic force microscopy. *Journal of Applied Physics* | bhushan |
| | Liu, H; **Bhushan**, B (2002). Investigation of nanotribological properties of self-assembled monolayers with alkyl and biphenyl spacer chains (Invited). *Ultramicroscopy* | liu |
| | Liu, HW; **Bhushan**, B (2003). Nanotribological characterization of molecularly thick lubricant films for applications to MEMS/NEMS by AFM. *Ultramicroscopy* | liu |
| | **Bhushan**, B (2003). Adhesion and stiction: Mechanisms, measurement techniques, and methods for reduction. *Journal of Vacuum Science and Technology B* | bhushan |
| | **Bhushan**, B (2005). Nanotribology and nanomechanics. *Wear* | bhushan |
| | Kim, SH; Asay, DB; Dugger, MT (2007). Nanotribology and MEMS. *Nano Today* | kim |
| Green | Tsukruk, VV (2001). Molecular lubricants and glues for micro- and nanodevices. *Advanced Materials* | tsukruk |
| | Eapen, KC; Patton, ST; Zabinski, JS (2002). Lubrication of microelectromechanical systems (MEMS) using bound and mobile phases of FomblinZdol (R). *Tribology Letters* | eapen |
| | Liu, W; Ye, C; Chen, Y; Ou, Z; Sun, DC (2002). Tribological behavior of sialon ceramics sliding against steel lubricated by fluorine-containing oils. *Tribology International* | liu |
| | Phillips, BS; Zabinski, JS (2004). Ionic liquid lubrication effects on ceramics in a water environment. *Tribology Letters* | phillips |
| | Lu, QM; Wang, HZ; Ye, CF; Liu, WM; Xue, QJ. (2004). Room temperature ionic liquid 1-ethyl-3-hexylimidazolium-bis(trifluoromethylsulfonyl)-imide as lubricant for steel-steel contact. *Tribology International* | lu |
| | Satyanarayana, N; Sinha, SK (2005). Tribology of PFPE overcoated self-assembled monolayers deposited on Si surface. *Journal of Physics D – Applied Physics* | satyanarayana |
| | Mu, ZG; Zhou, F; Zhang, SX; Liang, YM; Liu, WM (2005). Effect of the functional groups in ionic liquid molecules on the friction and wear behavior of aluminum alloy in lubricated aluminum-on-steel contact. *Tribology International* | mu |
| | Satyanarayana, N; Sinha, SK; Ong, BH (2006). Tribology of a novel UHMWPE/PFPE dual-film coated onto Si surface. *Sensors and Actuators A – Physical* | satyanarayana |
| | Jimenez, AE; Bermudez, MD; Iglesias, P; Carrion, FJ; Martinez-Nicolas, G (2006). 1-N-alkyl-3-methylimidazolium ionic liquids as neat lubricants and lubricant additives in steel-aluminium contacts. *Wear* | jimenez |
| | Jin, CM; Ye, CF; Phillips, BS; Zabinski, JS; Liu, XQ; Liu, WM; Shreeve, JM (2006). Polyethylene glycol functionalized dicationic ionic liquids with alkyl or polyfluoroalkyl substituents as high temperature lubricants. *Journal of Materials Chemistry* | jin |
| | Liu, XQ; Zhou, F; Liang, YM; Liu, WM (2006). Tribological performance of phosphonium based ionic liquids for an aluminum-on-steel system and opinions on lubrication mechanism. *Wear* | liu |
| | Qu, J; Truhan, JJ; Dai, S; Luo, H; Blau, PJ (2006). Ionic liquids with ammonium cations as lubricants or additives. *Tribology Letters* | qu |
| | Xia, YQ; Sasaki, S; Murakami, T; Nakano, M; Shi, L; Wang, HZ (2007). Ionic liquid lubrication of electrodeposited nickel-Si3N4 composite coatings. *Wear* | xia |
| | Jimenez, AE; Bermudez, MD (2007). Ionic liquids as lubricants for steel-aluminum contacts at low and elevated temperatures. *Tribology Letters* | jimenez |
| | Minami, I (2009). Ionic Liquids in Tribology. *Molecules* | minami |
| | Zhou, F; Liang, YM; Liu, WM (2009). Ionic liquid lubricants: designed chemistry for engineering applications. *Chemical Society Reviews* | zhou |
| | Bermudez, MD; Jimenez, AE; Sanes, J; Carrion, FJ (2009). Ionic Liquids as Advanced Lubricant Fluids. *Molecules* | bermudez |
| Brown | Dumitru, G; Romano, V; Weber, HP; Haefke, H; Gerbig, Y; Pfluger, E (2000). Laser microstructuring of steel surfaces for tribological applications. *Applied Physics A – Materials Science and Processing* | dumitru |
| | Pettersson, U; Jacobson, S (2003). Influence of surface texture on boundary lubricated sliding contacts. *Tribology International* | pettersson |
| | Donnet, C; Erdemir, A (2004). Solid lubricant coatings: recent developments and future trends. *Tribology Letters* | donnet |

## CONCLUSION

Tribology is a multidisciplinary engineering field in which the entire world is focusing on to reduce the consequences of friction and wear to increase the service life of industrial components. So, it is necessary to have a tool like CitNetExplorer to analyze the most cited publications that were published in the past to fine-tune the present research work. In this study, we examined the citation relations of publications on tribology for a period of 15 years from 1998 to 2012 based on WoS data using the software tool CitNetExplorer. With this software, we analyzed and clustered core publications. Also we studied the core publications of tribology (Figure 1), the publications on the topic – nanotribology (Figures 2 and 3), and the publications of a single researcher - namely Bharat Bhushan (Figure 4). For an easy understanding of the visualizations, the bibliographic details of the publications in the networks have been given in separate tables. Based on our experiences with CitNetExplorer in this study, we conclude that the software tool has the following advantages:

- Influential papers can be identified either on a topic or of a researcher.
- A particular paper can be visualized along with citing and cited publications.
- Common properties between publications can be identified through clustering.
- Drilling down facilities can be used to get a sub-network (e.g. from tribology to nanotribology publications).

We have the following three suggestions to improve the CitNetExplorer:

1. We studied the citation relations of publications on nanotribology. For this purpose, we used a set of keywords in the titles field to search the publications. In doing so, the following papers (partial list) have not been identified by the software:
   - Spin friction observed on the atomic scale
   - Friction-formed liquid droplets
   - Probe-tip induced damage in compliant substrates
   - On the application of transition state theory to atomic-scale wear

   In these papers, the keyword *nanotribology* has been used in abstracts. But the software searches the keywords in titles only. Hence, we suggest that the abstract and author keyword fields are also considered in a search besides the title.
2. There is no export facility of bibliographic records which are visualized in a map (here: the 40 most cited publications). In the visualization, one can get only information about the author, the article title, the source title, and the publication year of a particular paper.

3. CitNetExplorer uses internal citations only. If it provides external citations too, a user would get additional information about a particular paper.